\documentclass{icrc2009}

\usepackage{graphicx}   
\usepackage[caption=false]{caption}    
\usepackage[font=footnotesize]{subfig} 
\usepackage{fixltx2e}
\usepackage{url}

\newcommand{\shorttitle}[1]%
{\markboth{Proceedings of the 31\MakeLowercase{$^{st}$} ICRC, {\L}\'{o}d\'{z}
2009}{#1} }
\newcommand{\etal}{\MakeLowercase{\textit{et al. }}} 

\begin{document}
\title{{\it Fermi}-LAT Observation of Quiescent Solar Emission}

\author{\IEEEauthorblockN{Elena Orlando\IEEEauthorrefmark{1} on behalf of the {\it Fermi}-LAT
Collaboration
                          }
                            \\
\IEEEauthorblockA{\IEEEauthorrefmark{1} Max-Planck-Institut f\"ur extraterrestrische Physik, Postfach 1312, D-85741 Garching, Germany 
     }
}

\shorttitle{E.Orlando \etal Fermi-LAT Observation of Quiescent Solar Emission}
\maketitle

\begin{abstract}
The Large Area Telescope (LAT), one of two instruments on the {\it Fermi} Gamma-ray Space Telescope is a pair-conversion detector designed to study the gamma-ray sky in the energy range 30 MeV to 300 GeV. {\it Fermi} has detected high-energy gamma rays from the quiet Sun produced by interactions of cosmic-ray nucleons with the solar surface, and cosmic-ray electrons with solar photons in the heliosphere. While the Sun was detected by EGRET on CGRO with low statistics, {\it Fermi} provides high-quality detections on a daily basis allowing variability to be addressed. Such observations will provide a probe of the extreme conditions near the solar surface and a monitor the modulation of cosmic rays over the inner heliosphere. We discuss the study of the quiescent solar emission including spectral analysis of its two components, disk and inverse Compton.
\end{abstract}

 
\section{Introduction}
{\it Fermi} was successfully launched from Cape Canaveral on the 11th of June 2008.
 It is currently in a circular orbit around the Earth at
 an altitude of 565~km having an inclination of 25.6$^\circ$ and an orbital period
of 96 minutes. After an initial period of engineering data taking, the observatory
was  put into a sky-survey mode.
The observatory has two instruments onboard,
the Large Area Telescope (LAT)\cite{LATpaper},  a pair-conversion gamma-ray detector
and tracker   
  and a Gamma Ray Burst Monitor (GBM), dedicated to the detection of gamma-ray
	bursts. The instruments  on {\it Fermi} provide coverage over  the energy range
 from few kev to several hundred of GeV. 

The solar disk is a steady source of gamma rays produced by hadronic interactions of cosmic rays (CRs) in the solar atmosphere, the disk emission \cite{seckel1991}. The estimated disk flux of $~$10$^{-7}$ cm$^{-2}$s$^{-1}$ above 100 MeV from pion decays was at the limit of {\it EGRET} sensitivity \cite{thompson1997}.  Moreover the Sun was predicted to be an extended source of gamma-ray emission, produced by inverse Compton scattering of cosmic-ray electrons off solar photons in \cite{moska2006} and \cite{orlando2007}. This emission has a broad distribution on the sky with maximum intensity in the direction of the Sun. 
A detailed analysis of the {\it EGRET} data \cite{orlando2008} yielded a total flux of (4.44$\pm$2.03) $\times$ 10$^{-7}$ cm$^{-2}$s$^{-1}$ for E$>$100 MeV for these two solar components at 4 sigma, consistent with the predicted level. 
Both emission mechanisms have their maximum flux at the solar minimum condition when the cosmic-ray flux is maximum because of the lower level of solar modulation. 

Observations with {\it Fermi}-LAT of the inverse Compton scattered solar photons will allow for continuous monitoring of the cosmic-ray electron spectrum even in the close proximity of the solar surface; this is important for heliospheric cosmic-ray modulation studies. 
We report on the first months of observations and preliminary estimates of
observed fluxes and spectra.

\section{Data selection}
\begin{figure*}[h!]
   \centering-
\includegraphics[height=4in,width=5in]{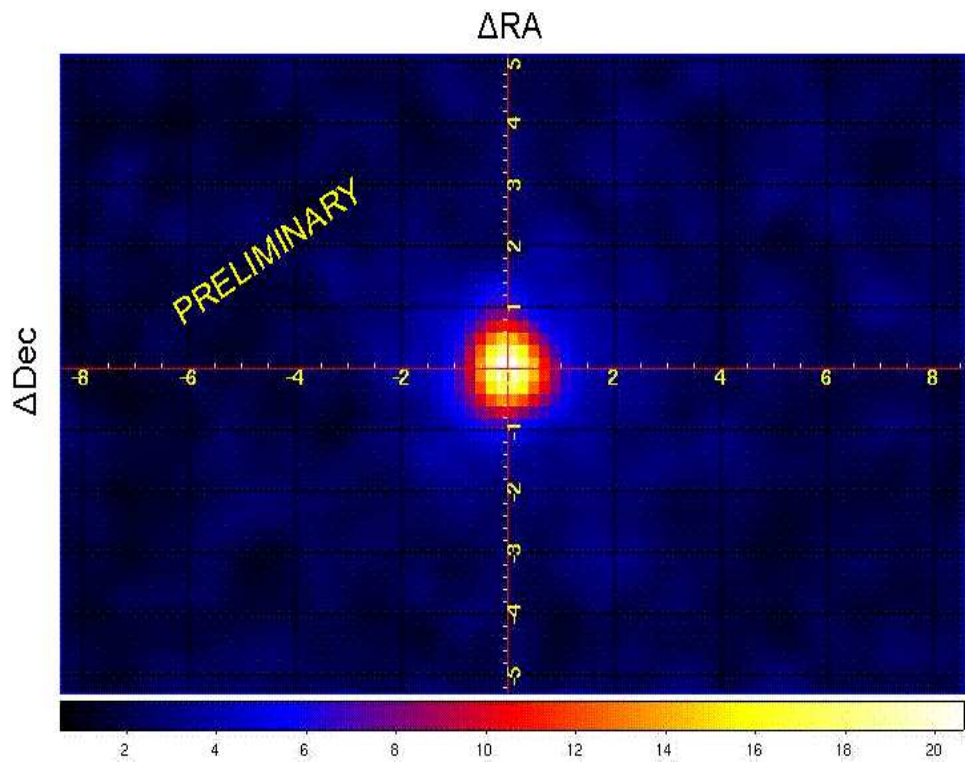}
	\caption{Count map in Sun-centred system for E $>$ 100MeV obtained with {\it Fermi}-LAT using data between August 2008 and end of January 2009. The pixel size used is 0.25$^{\circ}$ and the units are degrees on the axes and counts per pixel on the colour bar. The image is 2 bin radius Gaussian smoothed.}
  \label{fig1}
 
\end{figure*}

During the first year of the {\it Fermi}-LAT mission, the Sun was at its minimum in activity prior to the beginning of Solar Cycle 24.
The quiescent solar gamma-ray flux during this time period is therefore expected to be at its maximum. LAT is able to detect the solar emission almost daily when the Sun is not close to the Galactic plane and brightest sources. Figure~\ref{fig1} shows the count map above 100 MeV in Sun-centred coordinates integrated for a period from August 2008 and end of January 2009.  A significant photon excess corresponding to the position of the Sun is clearly visible.
This image is an update of the count map published in \cite{giglietto} with data accumulated from July 2008 to the end of September 2008 and in \cite{monica} with data collected during the first 6 months of the mission, from
August 2008 to January 2009. 
A preliminary analysis of the solar emission, using the standard unbinned maximum likelihood LAT
science tool was
performed by fitting a model for the
background and a point source with a simple power law for the Sun. (This is an approximated method since a more precise analysis should take into account the two different components of the solar emission, as described in the next Section).
This first analysis gives a detection with a test-statistics value of 2364 (defined as
TS = -2(log L$_0$-log L), where L and L$_0$ are the likelihood
values with and without the source respectively).

In order to remove contamination from the expected lunar emission, we exclude the data when the angular separation of the Moon and Sun is less than 30 degrees. Moreover, to avoid gamma-rays emitted from the Earth's atmosphere we only accepted events at zenith angles less than 105$^{\circ}$. For this analysis we use the "Diffuse" class
\cite{LATpaper}, corresponding to events with the highest probability of being photons. We also use Science Tools version v9r11 and IRFs (Instrumental Response
Functions) version P6\_V3 \footnote{http://fermi.gsfc.nasa.gov/ssc/data/analysis/documentation/Cicerone/ Cicerone\_LAT\_IRFs/IRF\_PSF.html}.
We remove the brightest celestial sources and the Galactic plane by excluding data when they are within few degrees of the Sun.  Moreover we use data covering uniform solar activity because of its low activity in the first year of {\it Fermi}-LAT observations.

\section{Analysis method}
As the Sun is a moving source we use software developed within the collaboration to analyze the data in a Sun-centred coordinate system.  We also use a likelihood fitting technique in our analysis, taking advantage of the previous experience gained with {\it EGRET} \cite{orlando2008}. 
We utilize a unique technique, the ``fake'' source method, to remove the anisotropic background component that includes diffuse Galactic and extragalactic emission, the instrumental background and sources not excluded from the data.   This method follows a source moving along the same ecliptic path as the Sun, but at a different time, to remove this residual background. We perform the same analysis of this ``fake Sun'' data as we do on the Sun, i.e.  a free model template containing both inverse Compton emission and  disk emission components.

\section{Results}
In our presentation we will include results from analysis of {\it Fermi} data collected during the first year of the mission, from August 2008 to June 2009. 
We will discuss spectral analysis of the two components (disk and extended inverse Compton) of the emission, and will compare the data with with models for different levels of solar modulation.
 
Models that provide the best fits to the data will be discussed and constraints on the electron spectrum will be placed. We will also discuss theoretical models and formulations of the modulation of cosmic rays in the inner heliosphere for solar minimum conditions. In particular we shall study    the extended inverse Compton emission at high energy, where the effect of the instruments PSF (point spread function) is small, in order to compare the spatial distribution with theoretical models.

\section{Acknowledgements}
The {\it Fermi}-LAT Collaboration acknowledges the generous support of a number of
agencies and institutes that have supported the {\it Fermi}-LAT Collaboration. These
include the National Aeronautics and Space Administration and the Department of
Energy in the United States, the Commissariat \`a l'Energie Atomique and the Centre
National de la Recherche Scientifique / Institut National de Physique Nucl\'eaire et
de Physique des Particules in France, the Agenzia Spaziale Italiana and the Istituto
Nazionale di Fisica Nucleare in Italy, the Ministry of Education, Culture, Sports,
Science and Technology (MEXT), High Energy Accelerator Research Organization (KEK)
and Japan Aerospace Exploration Agency (JAXA) in Japan, and the K.\ A.\ Wallenberg
Foundation and the Swedish National Space Board in Sweden.

\end{document}